# Tunable bilayer Hubbard model physics in twisted WSe$_2$

Yang Xu[1,2*], Kaifei Kang[1], Kenji Watanabe[3], Takashi Taniguchi[4], Kin Fai Mak[1,5,6*], and Jie Shan[1,5,6*]

[1]School of Applied and Engineering Physics, Cornell University, Ithaca, NY, USA.
[2]Beijing National Laboratory for Condensed Matter Physics, Institute of Physics, Chinese Academy of Sciences, Beijing, China.
[3]Research Center for Functional Materials, National Institute for Materials Science, Tsukuba, Japan.
[4]International Center for Materials Nanoarchitectonics, National Institute for Materials Science, Tsukuba, Japan.
[5]Department of Physics, Cornell University, Ithaca, NY, USA.
[6]Kavli Institute at Cornell for Nanoscale Science, Ithaca, NY, USA.

*Email: yang.xu@iphy.ac.cn; kinfai.mak@cornell.edu; jie.shan@cornell.edu

**Moiré materials with flat electronic bands provide a highly controllable quantum system for studies of strong-correlation physics and topology [1-3]. In particular, angle-aligned heterobilayers of semiconducting transition metal dichalcogenides (TMDs) with large band offset realize the single-band Hubbard models [4-11]. Introduction of a new layer degree of freedom is expected to foster richer interactions, enabling Hund's physics, interlayer exciton condensation and new superconducting pairing mechanisms to name a few [12-15]. Here, we report competing electronic orders in twisted AB-homobilayer WSe$_2$, which realizes a bilayer Hubbard model in the weak interlayer hopping limit for holes [15, 16]. We characterize the charge order, layer polarization and magnetization of the moiré bilayer, subjected to an out-of-plane electric and magnetic field, by exciton sensing and magneto circular dichroism measurements. By layer-polarizing holes via the electric field, we observe a crossover from an excitonic insulator to a charge-transfer insulator at hole density of $\nu = 1$ (in unit of moiré density), a transition from a paramagnetic to an antiferromagnetic charge-transfer insulator at $\nu = 2$, and evidence for a layer-selective Mott insulator at $1 < \nu < 2$. The unique coupling of charge and spin to external fields also manifests a giant magneto-electric response. Our results establish a new solid-state simulator for problems in strong-correlation physics that are described by bilayer Hubbard models [12-15].**

**Main**

The interlayer coupling in TMD homobilayers is strongly dependent on the twist angle between the layers [16-23]. In the vicinity of the valence band edge located at the K and -K points of the Brillouin zone, the bands of each monolayer are spin-polarized with large spin splitting, and the spin and valley are locked. For near-0°-twisted (or AA-stacked) homobilayers, the layers are strongly hybridized [16, 17, 19-23]. Experimental studies demonstrate a correlated insulating state at doping density of one hole per moiré unit cell [24-26]. It is argued that the single-band Hubbard model well approximates the system. On the other hand, for a natural homobilayer, which is 60°-twisted (or AB-stacked), the single-particle interlayer tunneling is spin forbidden [15-17, 19, 20]. In this limit, the layer degree of freedom can be considered as layer pseudospin; together with spin it forms SU(4)-spin (Fig. 1c, d). Although recent theoretical studies predict SU(4) chiral spin liquid and exciton supersolid phases in twisted TMD AB-homobilayers [15], experimental studies of the charge, layer and spin correlations remain elusive.

Here, we investigate twisted $WSe_2$ AB-homobilayers as a prototype bilayer Hubbard system. The twist angle is kept within 2 - 3° of AB stacking to maintain flat bands for strong correlation effects. We demonstrate results from a 58°-twisted bilayer. It forms a triangular moiré lattice with period of about 10 nm and moiré density of $n_M \approx 1.1 \times 10^{12}$ cm$^{-2}$. Scanning tunneling microscopy and spectroscopy reveal that the first moiré flat band for holes is localized on the MX site (M = W and X = Se) in the moiré unit cell [18] (Fig. 1a). The sample is encapsulated in two nearly symmetric gates that are made of hexagonal boron nitride (hBN) gate dielectrics and graphite gate electrodes; it is grounded through a graphite contact (Fig. 1b). The two gates allow independent tuning of the doping density ($\nu$) and perpendicular electric field ($E$) in the sample.

A local dielectric sensor is integrated into the right half of the device. It consists of a $WSe_2$ monolayer that is electronically decoupled from the sample by a ~ 1-nm-thick hBN spacer. The sensor's excited exciton resonances are highly sensitive to the screening effect [7, 27, 28], either from the emergence of correlated insulating states in the nearby sample or a change in the sample-sensor separation. The latter provides access to the sample layer polarization ($P = \frac{\nu_t - \nu_b}{\nu_t + \nu_b}$ with $\nu_t$ and $\nu_b$ denoting the charge density on the top and bottom layers of the moiré bilayer, respectively). We also probe the sample magnetization ($M$) and magnetic susceptibility ($\chi$) by magneto circular dichroism (MCD) spectroscopy of the moiré exciton under an out-of-plane magnetic field $B$. Details on the device fabrication and optical measurements are provided in Methods. All results are obtained at 1.7 K unless otherwise specified.

**Insulating states**
Figure 1e shows the reflectance contrast spectrum of the right half of the device as a function of gate voltage (bottom axis) and doping density (top axis) in the moiré bilayer. The sensor layer is kept charge neutral. The resonance feature near 1.725 eV that saturates the color scale and the feature near 1.85 eV are, respectively, the 1s and 2s excitons of the $WSe_2$ monolayer sensor. The feature that is red-shifted from the sensor 1s exciton by 30 - 50 meV is the moiré exciton, whose energy is lowered by the interlayer coupling in the $WSe_2$ moiré bilayer. The reflectance contrast spectrum of the left half of the device is similar, minus the sensor features.

A series of insulating states are revealed by the sensor 2s exciton as abrupt energy blueshift, correlated with reflectance contrast enhancement [7]. The strongest insulating state is observed for the charge-neutral sample ($\nu = 0$). Upon electron doping, additional insulating states emerge at $\nu = 1, 2, 3, 4$, and several fractional filling factors. Upon hole doping, only the $\nu = 1$ and 2 insulating states are clearly visible. This suggests flatter moiré conduction bands since many of these states arise from the correlation effect. In addition, the insulating states for the electrons are not sensitive to out-of-plane electric field $E$, indicating strong interlayer hybridization of the states (which are from the Q points of the Brillouin zone [29]). In contrast, the hole states strongly depend on $E$ (see below), indicating weak interlayer hybridization. We will focus on the case of hole doping to investigate the correlation effects in the weak interlayer coupling limit.

**Electric-field control of layer polarization**
We examine the electric-field effect on layer polarization at a fixed hole density in the right region of the moiré bilayer. Figure 2a illustrates the sensor 2s spectrum at $\nu = 1$. The 2s exciton is discernable for $E < 20$ mV/nm, above which the sensor is doped as a result of charge transfer from the moiré bilayer and the 2s exciton is quenched (Extended Data Fig. 1,2). As field sweeps from 20 to -30 mV/nm, the 2s exciton exhibits a blueshift over a small field range around 0 mV/nm; the response is flat outside this range. The blueshift signals reduced screening of the sensor exciton when charge transfers from the bottom to the top layer in the moiré bilayer (i.e. further away from the sensor). Above the critical fields (white dashed lines), the charges are fully layer-polarized ($P = \pm 1$); below, they are shared between the layers. We summarize the doping dependence of the critical field, $E_c$, in Fig. 2c. For large doping densities, only the crossover to the $P = +1$ state is accessible because doping into the sensor limits its operating electric-field range.

A complementary moiré exciton probe supports the interpretation. Figure 2b shows the field dependent moiré exciton resonance measured at $\nu = 2$ from the left region of the device (see Extended Data Fig. 5 for other densities; the twist angle in the two regions differs by about 0.1°). The spectrum shows two distinct features (guided by the black dashed lines) under small $E$'s and a single prominent feature above $|E_c|$ (white dashed lines). Similar effect is observed in MoSe$_2$ bilayers separated by a thin hBN layer [25]. The single prominent feature is interpreted as the strong response of the charge-neutral layer after the bilayer is fully layer-polarized. The two spectral features under small $E$'s presumably arise from layer asymmetry and/or a small interlayer mixing when both layers are doped.

We map the layer polarization as a function of ($\nu, E$) in Fig. 2d by tracing the reflectance contrast averaged around the single prominent spectral feature ($\approx 1.7$ eV). Regions with low (high) reflectance contrast correspond to partial (full) layer polarization. The exact value for $|P| < 1$ cannot be calibrated from this measurement. The result is fully consistent with Fig. 2c. The threshold field increases linearly with $\nu$ in each of the three doping regions, $\nu < 1$, $1 < \nu < 2$ and $\nu > 2$; and the slope increases from the low to high density regions. There is also a notable jump in $E_c$ at $\nu = 1$ and 2, reflecting the charge gap of the states. Intriguingly, the observed $E_c$ for $\nu < 1$ is more than an order of magnitude smaller than the estimate from electrostatics using the parallel plate capacitor model (Methods). This reflects strong correlation effects that favor layer-polarized states. However, ferroelectric instability (including spontaneous layer polarization and electric-field hysteresis) is not observed.

**Magnetic properties**
To gain more insight into the nature of the electronic states, we probe their magnetic properties as a function of ($\nu, E$) by MCD. The MCD spectrum of the left region of the device is averaged over a narrow spectral window (typically 5 meV) centered on the moiré exciton (Extended Data Fig. 6); the value is linearly proportional to magnetization [6, 11]. Figure 3a-c shows the magnetic-field dependence of MCD at three representative doping densities, $\nu = 1$, 1.6 and 2. The top and bottom panels correspond to the case of $P = 0$ and 1, respectively. For layer-unpolarized holes at $\nu = 1$, we observe a paramagnetic (PM) response at all temperatures, that is, the MCD increases from zero as $B$ increases, but the rate decreases monotonically. We extract the magnetic susceptibility, $\chi$, from the MCD slope near zero magnetic field (Fig. 3d). It follows the Curie-Weiss law, $\chi^{-1} \propto T - \theta$, with Weiss temperature $\theta \approx -1$ K (blue line). This signals the emergence of magnetic local moments. The negative Weiss temperature reflects antiferromagnetic (AF) exchange interaction between the local moments. In addition, the MCD fully saturates at ~ 2 T at low temperature. Similar behavior is observed for full layer polarization with slightly weaker AF exchange interaction (black symbols, Fig. 3d).

For layer-unpolarized holes at $\nu = 2$ (Fig. 3c), the magnetic response is similar to that at $\nu = 1$. The Weiss temperature is about -2 K; the MCD fully saturates at ~2 T at low temperature. In contrast, for layer-polarized holes the magnetic response is PM only at temperatures above ~10 K. Below 10 K, the response is metamagnetic, that is, the MCD increases slowly with $B$ for small fields, followed by a faster increase before reaching saturation at ~ 6 T. The high-temperature magnetic susceptibility follows the Curie-Weiss law with $\theta \approx -9$ K. Furthermore, $\chi$ shows a broad peak near $|\theta|$. These behaviors support AF ordering below 9 K (Fig. 3f).

For $1 < \nu < 2$ (Fig. 3b,e), the magnetic response is again PM for layer-unpolarized holes. It develops two components for layer-polarized holes: PM at small magnetic fields and metamagnetic at larger magnetic fields. Extended Data Fig. 8 illustrates the evolution as a function of electric field; the metamagnetic component appears only above $E_c$ (i.e. reaching full layer polarization). In addition, as a function of doping density the PM component continuously shrinks to zero when $\nu$ approaches 2 (Extended Data Fig. 7). The saturation magnetic field for the metamagnetic component increases slowly with density.

**Nature of the correlated electronic states**

Our experiment shows that the $\nu = 1$ and 2 states remain insulating for both layer-polarized and -unpolarized charges (Extended Data Fig. 3), under the largest magnetic field accessible in this experiment (9 T) and up to about 80 K (Extended Data Fig. 4). Below we discuss the nature of their ground states and illustrate the most plausible charge/spin configuration in the insets of Fig. 3a-c.

The measurement shows that the $\nu = 1$ state with $|P| = 1$ is a PM insulator down to 1.7 K. In the limit of weak interlayer coupling, the situation is equivalent to half-filling in a TMD moiré heterobilayer with large band offset [6]. The state is a Mott insulator or a charge-transfer insulator in the presence of more than one orbital (or stacking site with moiré potential minimum [8]) as in this case. The holes are localized by the strong on-site Coulomb repulsion on the MX site of the moiré lattice in one of the WSe$_2$ layers (Fig. 3a) [18]. In the flat band limit, the local moments interact via the AF super-exchange mechanism [6] ($\theta < 0$), which is weak for distant moments (small $|\theta|$). The state thus remains PM down to 1.7 K; and a relatively small magnetic field (~ 2 T) is sufficient to align these 'nearly isolated' spins at low temperature. With $|P| < 1$, the holes are distributed in both layers. The holes in one layer are bound to the empty moiré sites in the other layer to minimize the (interlayer) on-site Coulomb repulsion. Under a particle-hole transformation in the Hubbard band when the spins are ignored, the empty sites become electrons; the state is an excitonic insulator. Such a phase is theoretically proposed for twisted TMD AB-bilayers [15] and is observed in a related system of Coulomb coupled TMD monolayer-moiré bilayer heterostructures [30, 31].

Our MCD results show that the $\nu = 2$ state with $|P| = 1$ is an AF-ordered insulator below 9 K. The state is not compatible with a simple band insulator, in which two holes with antiparallel spins occupy the same orbital. In such an insulator, the spin gap would equal the charge gap following the Pauli exclusion principle. Contrarily, here the charge gap, estimated from twice of the gap-closure temperature of 80 K, is substantially larger than the spin gap, estimated from the Weiss temperature of 9 K. The conclusion is further supported by the robust insulating state observed beyond magnetic saturation (Extended Data Fig. 4). The most natural scenario is therefore a charge-transfer insulator with a Néel-type AF order [8]. Two holes with antiparallel spins occupy two different stacking sites (likely MX and MM) in each moiré unit cell that form a honeycomb lattice (Fig. 3c). The closer proximity of local moments in this configuration explains the enhanced $|\theta|$, and a much higher magnetic field (~6 T) to overcome the exchange interaction to fully polarize the spins. The removal of geometrical frustration associated with triangular lattices also helps to stabilize AF ordering.

On the other hand, the insulating state at $\nu = 2$ with $P \approx 0$ is practically two copies of the $\nu = 1$ state with $|P| = 1$. It is a charge-transfer insulator; the holes in two different layers are likely localized on two different stacking sites of the triangular lattice to minimize the (interlayer) on-site Coulomb repulsion. The transition from the $|P| = 0$ to 1 state is rather abrupt near $E_c$ (see below).

Whenever the density in a monolayer exceeds 1 for $1 < \nu < 2$, the excess spins prefer to anti-align with their neighbors because of the enhanced AF exchange interaction; they form 'AF clusters' in a honeycomb lattice (Fig. 3b). The presence of both nearly isolated moments and AF clusters gives rise to the observed two-component magnetic responses for $1 < \nu < 2$. The composition of the PM and metamagnetic responses is electric-field tunable. This picture is clearly illustrated in the ($\nu$, $E$)-dependence of the MCD at $B = 2$ T (Fig. 4a). At 2 T, the nearly isolated moments are fully saturated, but the AF clusters remain largely unperturbed and their contribution to the MCD is small. The MCD thus mostly reflects the density of isolated moments. Figure 4b presents two extreme cases with $|P| = 0$ and 1. For full layer polarization (black symbols), the density of isolated moments increases linearly with $\nu$ from zero till reaching $\nu = 1$; it then decreases linearly for $\nu > 1$ because the nearly isolated moments are continuously replaced by the AF clusters; the nearly isolated moments are completely exhausted at $\nu = 2$. On the other hand, for $P = 0$ (blue symbols) the same dependence appears at twice of the density because now both layers are involved.

The unprecedented gate control of the competing electronic phases in twisted TMD AB-bilayers opens up a new way to realize strong correlation effects and applications. In particular, because of the robust Mott gap and the flat Hubbard bands in each strongly correlated monolayer, a layer-selective Mott insulator [32], in which one layer is a Mott (or charge-transfer) insulator and the other layer contains itinerant electrons, can be achieved by adjusting the electric field for $v > 1$. The sudden charge transfer near $E_c$ (Extended Data Fig. 2,5) supports the emergence of a layer-selective Mott insulator for $1 < v < 2$. This configuration with metallic states interacting with a lattice of local moments can potentially be used to study the Kondo lattice model [32, 33]. Furthermore, an electric field can switch the system at $v = 2$ from a layer-unpolarized PM insulator to a layer-polarized AF insulator, which results in a drastic change in magnetization under a finite magnetic field (Fig. 4, Methods). This effectively is a giant magneto-electric (ME) effect (i.e. control of the magnetization by an electric field and vice versa), and is of a purely electronic origin. The electronic ME effect [34] is sought-after for energy-efficient, high-speed sensing and information technology, but has remained elusive in conventional solid-state materials [35]. With a variety of competing electronic orders driven by strong correlation and with charge and spin weakly coupled to the lattice, moiré materials will likely shed new light on electronic magnetoelectrics.

**Methods**
**Device fabrication.** Flakes of few-layer graphite, hBN, and monolayer $WSe_2$ are mechanically exfoliated from bulk crystals on Si substrates with a 285-nm $SiO_2$ layer. They are selected by the color contrast under an optical microscope. The top and back hBN gate dielectrics are chosen to have nearly identical thickness (~ 40 nm). The heterostructure is assembled layer-by-layer using a dry transfer method described elsewhere [7]. The twisted $WSe_2$ bilayer is fabricated by the 'tear-and-stack' method, similar to that developed for twisted bilayer graphene [36-38]. Namely, part of a large $WSe_2$ monolayer is picked up and stacked on the remaining part with twist angle $\delta$ that is precisely controlled by a rotary stage. The target angle is 58°. The twist angle variation is about 0.1° across the sample, which is calibrated by the density of the insulating states at integer fillings locally. An additional $WSe_2$ monolayer is introduced into one part of the device as a sensor. It is separated from the twisted bilayer by a thin hBN layer (1-2 nm).

**Optical measurements.** The optical measurements are performed in a closed-cycle cryostat (Attocube, Attodry 2100) at temperatures down to 1.7 K and magnetic fields up to 9 T. The details of the reflectance spectrum measurement are described in Ref. [7]. Here the reflectance contrast, $\Delta R/R_0 = (R - R_0)/R_0$, is obtained by comparing reflectance $R$ to reference $R_0$ that is measured at the same sample location under large negative top and back gate voltages ($V_{tg} = V_{bg} = -8$ V). The magneto circular dichroism (MCD), $MCD = \frac{R_+ - R_-}{R_+ + R_-}$, is obtained using the reflectance spectrum excited by left and right circularly polarized light, $R_+$ and $R_-$, respectively. An example is shown in Extended Data Fig. 6. The reported MCD in this work is the average MCD over a small spectral range (typically 5 meV) centered near the moiré exciton resonance. To be more efficient, in some $B$-dependent studies we measure the reflectance spectrum with circularly polarized light of only one handedness (for example, $R_+$). We infer the spectrum of the other handedness based on $R_-(B) = R_+(-B)$.

Two regions on the device are studied. One is with the sensor layer; the twist angle is calibrated to be 58.2°. The sensor 2s exciton is utilized to detect the correlated insulating states and the layer polarization. The second region is without the sensor layer; the twist angle is 58.1°. The moiré exciton response and MCD are measured. The out-of-plane electric field is calculated from the gate voltages using $E = (C_{tg}V_{tg} - C_{bg}V_{bg})/2\epsilon_0\epsilon_{BN} - E_0$. Here $C_{tg}$ ($C_{bg}$) is the geometrical capacitance per unit area of the top (back) gate, $\epsilon_{BN} \approx 3$ is the dielectric constant of hBN, $\epsilon_0$ is the vacuum permittivity, and $E_0 \approx$ - 6 mV/nm is a small built-in electric field (likely arisen from small layer asymmetry of the structure). The latter is introduced so that the layer polarization is symmetric about $E = 0$. The top and back gate capacitances are individually calibrated by analyzing the gate dependence of the Landau levels in a WSe$_2$ monolayer that is attached to the moiré bilayer under a perpendicular magnetic field of 9 T. The Landau level is detected optically by measuring the polaron energy/intensity oscillation [28]. The two gate capacitances differ by less than 10%.

**Estimate of $E_c$ using electrostatics.** Charge transfer between the two WSe$_2$ monolayers in the twisted bilayer can be modeled by the parallel plate capacitor model. The critical electric field $E_c$ required to achieve full layer polarization for charge density $\nu n_M$ is

$$E_c = \frac{C_q + 2C_i + C_g}{C_q} \frac{\nu n_M e}{2\epsilon_{BN}\epsilon_0}. \tag{1}$$

Here $C_q$ is the quantum capacitance of each WSe$_2$ monolayer, $C_i$ is the interlayer capacitance of the twisted bilayer, $C_g = C_{tg}(\approx C_{bg})$ is the gate capacitance, and $e$ is the elementary charge. For non-interacting electrons, we have $C_q \gg C_i \gg C_g$ and $E_c \approx \frac{\nu n_M e}{2\epsilon_{BN}\epsilon_0} \approx 28$ mV/nm for $\nu = 1$. This value is about an order of magnitude larger than the observed $E_c$, which demonstrates the importance of the electron-electron interactions. The interactions significantly modify the value of $C_q$ and $C_i$ and favor the layer-polarized state. At $\nu = 2$, the discrepancy between experiment ($E_c \approx$ 17 mV/nm) and model ($E_c \approx$ 56 mV/nm) is reduced. In general, the discrepancy in $E_c$ is reduced with increasing doping density. This suggests that the interaction effects become weaker as the Fermi level is moved to higher-lying moiré bands. This is expected as the moiré bandwidth generally increases with the band index [39].

As discussed in the main text, there is a jump in $E_c$ at $\nu = 1$ and 2. The jump is caused by the presence of the charge gap for these insulating states. Additional electric field is required to overcome the charge gap and fully transfer the charges from one layer to the other [30]. However, the larger jump in $E_c$ observed at $\nu = 2$ than 1 is not well understood. It is correlated with the enhanced layer polarizability at low doping densities, at which the interaction effect is most important. Future theoretical studies are required to quantitatively describe the density dependence of the layer polarizability.

**Magneto-electric (ME) effect.** The coupling of the layer and spin degrees of freedom to the electric and magnetic fields, respectively, and the sensitivity of the magnetic ground state to layer polarization in twisted bilayer $WSe_2$ provide a possibility to achieve a large ME effect of pure electronic origin. We investigate the case at $\nu = 2$. Extended Data Fig. 9 shows the contour plot of the MCD as a function of ($E$, $B$). The MCD normalized by its value at 9 T (above magnetic saturation) characterizes the magnetization $M$ ($= \pm 1$ for a fully spin polarized state). Four different regions can be identified: (I) $|M|$, $|P| < 1$; (II) $|M| < 1$, $|P| = 1$; (III) $|M| = 1$, $|P| < 1$; and (IV) $|M|$, $|P| = 1$. The boundaries of these regions are determined according to $\partial M/\partial E$ and $\partial M/\partial B$. The magnetization is sensitive to both electric and magnetic fields. Particularly at intermediate magnetic fields, for $|P| = 1$ the strong intralayer AF exchange favors antiparallel spin alignments which reduce the magnetization substantially compared to the PM state for $P = 0$. An electric field can thus shift the system between region I and II and cause over 60% change in the sample magnetization. This is a giant ME effect. At high enough magnetic fields that overcome the intralayer AF exchange, the electric field shifts the system between region III and IV; the ME coupling diminishes. The ME effect is also doping dependent. Figure 4a shows magnetization as a function of electric field and doping density at a fixed magnetic field of 2 T. The magnetization changes significantly around the threshold electric field for filling $\nu > 1$. The ME effect is the largest around $\nu = 2$ and weakens for $1 < \nu < 2$ and $\nu > 2$.

**References**
1. Andrei, E.Y. & MacDonald, A.H. Graphene bilayers with a twist. *Nature Materials* **19**, 1265-1275 (2020).
2. Balents, L., Dean, C.R., Efetov, D.K. & Young, A.F. Superconductivity and strong correlations in moiré flat bands. *Nature Physics* **16**, 725-733 (2020).
3. Kennes, D.M. et al. Moiré heterostructures as a condensed-matter quantum simulator. *Nature Physics* **17**, 155-163 (2021).
4. Wu, F., Lovorn, T., Tutuc, E. & MacDonald, A.H. Hubbard Model Physics in Transition Metal Dichalcogenide Moiré Bands. *Physical Review Letters* **121**, 026402 (2018).
5. Regan, E.C. et al. Mott and generalized Wigner crystal states in $WSe_2/WS_2$ moiré superlattices. *Nature* **579**, 359-363 (2020).
6. Tang, Y. et al. Simulation of Hubbard model physics in $WSe_2/WS_2$ moiré superlattices. *Nature* **579**, 353-358 (2020).
7. Xu, Y. et al. Correlated insulating states at fractional fillings of moiré superlattices. *Nature* **587**, 214-218 (2020).
8. Zhang, Y., Yuan, N.F.Q. & Fu, L. Moiré quantum chemistry: Charge transfer in transition metal dichalcogenide superlattices. *Physical Review B* **102**, 201115 (2020).
9. Pan, H., Wu, F. & Das Sarma, S. Quantum phase diagram of a Moiré-Hubbard model. *Physical Review B* **102**, 201104 (2020).
10. Huang, X. et al. Correlated insulating states at fractional fillings of the $WS_2/WSe_2$ moiré lattice. *Nature Physics* **17**, 715-719 (2021).

11. Li, T. et al. Continuous Mott transition in semiconductor moiré superlattices. *Nature* **597**, 350-354 (2021).
12. dos Santos, R.R. Magnetism and pairing in Hubbard bilayers. *Physical Review B* **51**, 15540-15546 (1995).
13. Golor, M., Reckling, T., Classen, L., Scherer, M.M. & Wessel, S. Ground-state phase diagram of the half-filled bilayer Hubbard model. *Physical Review B* **90**, 195131 (2014).
14. Gall, M., Wurz, N., Samland, J., Chan, C.F. & Köhl, M. Competing magnetic orders in a bilayer Hubbard model with ultracold atoms. *Nature* **589**, 40-43 (2021).
15. Zhang, Y.-H., Sheng, D.N. & Vishwanath, A. SU(4) Chiral Spin Liquid, Exciton Supersolid, and Electric Detection in Moiré Bilayers. *Physical Review Letters* **127**, 247701 (2021).
16. Wu, F., Lovorn, T., Tutuc, E., Martin, I. & MacDonald, A.H. Topological Insulators in Twisted Transition Metal Dichalcogenide Homobilayers. *Physical Review Letters* **122**, 086402 (2019).
17. Naik, M.H. & Jain, M. Ultraflatbands and Shear Solitons in Moiré Patterns of Twisted Bilayer Transition Metal Dichalcogenides. *Physical Review Letters* **121**, 266401 (2018).
18. Zhang, Z. et al. Flat bands in twisted bilayer transition metal dichalcogenides. *Nature Physics* **16**, 1093-1096 (2020).
19. Pan, H., Wu, F. & Das Sarma, S. Band topology, Hubbard model, Heisenberg model, and Dzyaloshinskii-Moriya interaction in twisted bilayer WSe$_2$. *Physical Review Research* **2**, 033087 (2020).
20. Zhai, D. & Yao, W. Theory of tunable flux lattices in the homobilayer moiré of twisted and uniformly strained transition metal dichalcogenides. *Physical Review Materials* **4**, 094002 (2020).
21. Angeli, M. & MacDonald, A.H. Γ valley transition metal dichalcogenide moiré bands. *Proceedings of the National Academy of Sciences* **118**, e2021826118 (2021).
22. Devakul, T., Crépel, V., Zhang, Y. & Fu, L. Magic in twisted transition metal dichalcogenide bilayers. *Nature Communications* **12**, 6730 (2021).
23. Xian, L. et al. Realization of nearly dispersionless bands with strong orbital anisotropy from destructive interference in twisted bilayer MoS$_2$. *Nature Communications* **12**, 5644 (2021).
24. Wang, L. et al. Correlated electronic phases in twisted bilayer transition metal dichalcogenides. *Nature Materials* **19**, 861-866 (2020).
25. Shimazaki, Y. et al. Strongly correlated electrons and hybrid excitons in a moiré heterostructure. *Nature* **580**, 472-477 (2020).
26. Ghiotto, A. et al. Quantum criticality in twisted transition metal dichalcogenides. *Nature* **597**, 345-349 (2021).
27. Raja, A. et al. Coulomb engineering of the bandgap and excitons in two-dimensional materials. *Nature Communications* **8**, 15251 (2017).
28. Xu, Y. et al. Creation of moiré bands in a monolayer semiconductor by spatially periodic dielectric screening. *Nature Materials* **20**, 645-649 (2021).

29. Liu, G.-B., Xiao, D., Yao, Y., Xu, X. & Yao, W. Electronic structures and theoretical modelling of two-dimensional group-VIB transition metal dichalcogenides. *Chemical Society Reviews* **44**, 2643-2663 (2015).
30. Jie Gu, L.M., Song Liu, Kenji Watanabe, Takashi Taniguchi, James C. Hone, Jie Shan, Kin Fai Mak. Dipolar excitonic insulator in a moire lattice. *arXiv:2108.06588* (2021).
31. Zuocheng Zhang, E.C.R., Danqing Wang, Wenyu Zhao, Shaoxin Wang, Mohammed Sayyad, Kentaro Yumigeta, Kenji Watanabe, Takashi Taniguchi, Sefaattin Tongay, Michael Crommie, Alex Zettl, Michael P. Zaletel, Feng Wang. Correlated interlayer exciton insulator in double layers of monolayer $WSe_2$ and moiré $WS_2/WSe_2$. *arXiv:2108.07131* (2021).
32. Dalal, A. & Ruhman, J. Orbitally selective Mott phase in electron-doped twisted transition metal-dichalcogenides: A possible realization of the Kondo lattice model. *Physical Review Research* **3**, 043173 (2021).
33. Ajesh Kumar, N.C.H., Allan H. MacDonald, Andrew C. Potter. Gate-tunable heavy fermion quantum criticality in a moiré Kondo lattice. *arXiv:2110.11962* (2021).
34. Sivadas, N., Okamoto, S. & Xiao, D. Gate-Controllable Magneto-optic Kerr Effect in Layered Collinear Antiferromagnets. *Physical Review Letters* **117**, 267203 (2016).
35. Fiebig, M. Revival of the magnetoelectric effect. *Journal of Physics D: Applied Physics* **38**, R123-R152 (2005).
36. Kim, K. et al. van der Waals Heterostructures with High Accuracy Rotational Alignment. *Nano Letters* **16**, 1989-1995 (2016).
37. Cao, Y. et al. Superlattice-Induced Insulating States and Valley-Protected Orbits in Twisted Bilayer Graphene. *Physical Review Letters* **117**, 116804 (2016).
38. Chung, T.-F., Xu, Y. & Chen, Y.P. Transport measurements in twisted bilayer graphene: Electron-phonon coupling and Landau level crossing. *Physical Review B* **98**, 035425 (2018).
39. Li, T. et al. Charge-order-enhanced capacitance in semiconductor moiré superlattices. *Nature Nanotechnology* **16**, 1068-1072 (2021).

**Figures**

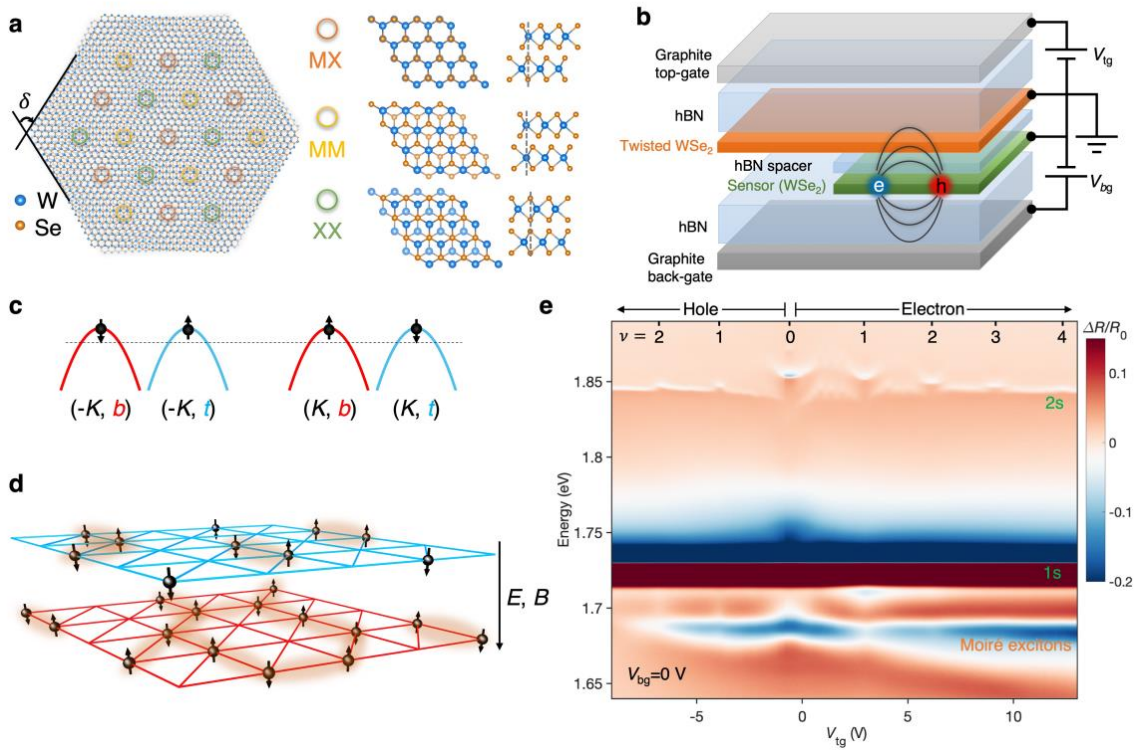

**Figure 1 | Twisted WSe$_2$ AB-homobilayer. a,** Twisted WSe$_2$ bilayer exhibits three types of high-symmetry stacking sites MX, MM and XX (M = W and X = Se) for twist angle $\delta \approx 60°$. **b,** Schematic of a dual-gated WSe$_2$ moiré bilayer device. The right half contains an exciton (bound *e-h* pair) sensor that is made of WSe$_2$ monolayer and separated from the sample by a thin hBN spacer. **c,** Hole spin alignment at the -K and K valleys in the top (*t*) and bottom (*b*) layers of WSe$_2$ AB-homobilayers. Dashed line denotes the Fermi level. Spin and valley are locked in each layer. Within each valley spins are anti-aligned in two layers. Interlayer tunneling is spin-forbidden. **d,** Illustration of the bilayer Hubbard model with intra-layer hopping much smaller than intra- and inter-layer onsite repulsions. The spin and layer degrees of freedom are addressable by an out-of-plane magnetic (*B*) and electric field (*E*), respectively. The arrow denotes the positive field directions. **e,** Reflectance contrast ($\Delta R/R_0$) spectrum of the device with the sensor versus top gate voltage. The back gate is fixed at 0 V. The corresponding doping density is shown on the top axis. The 2s and 1s exciton resonances of the sensor and the moiré exciton resonances of the twisted bilayer are shown in descending order of energy. A series of insulating states, manifested as blueshift of the sensor 2s resonance at integer and some fractional filling factors, are observed for both electron and hole doping.

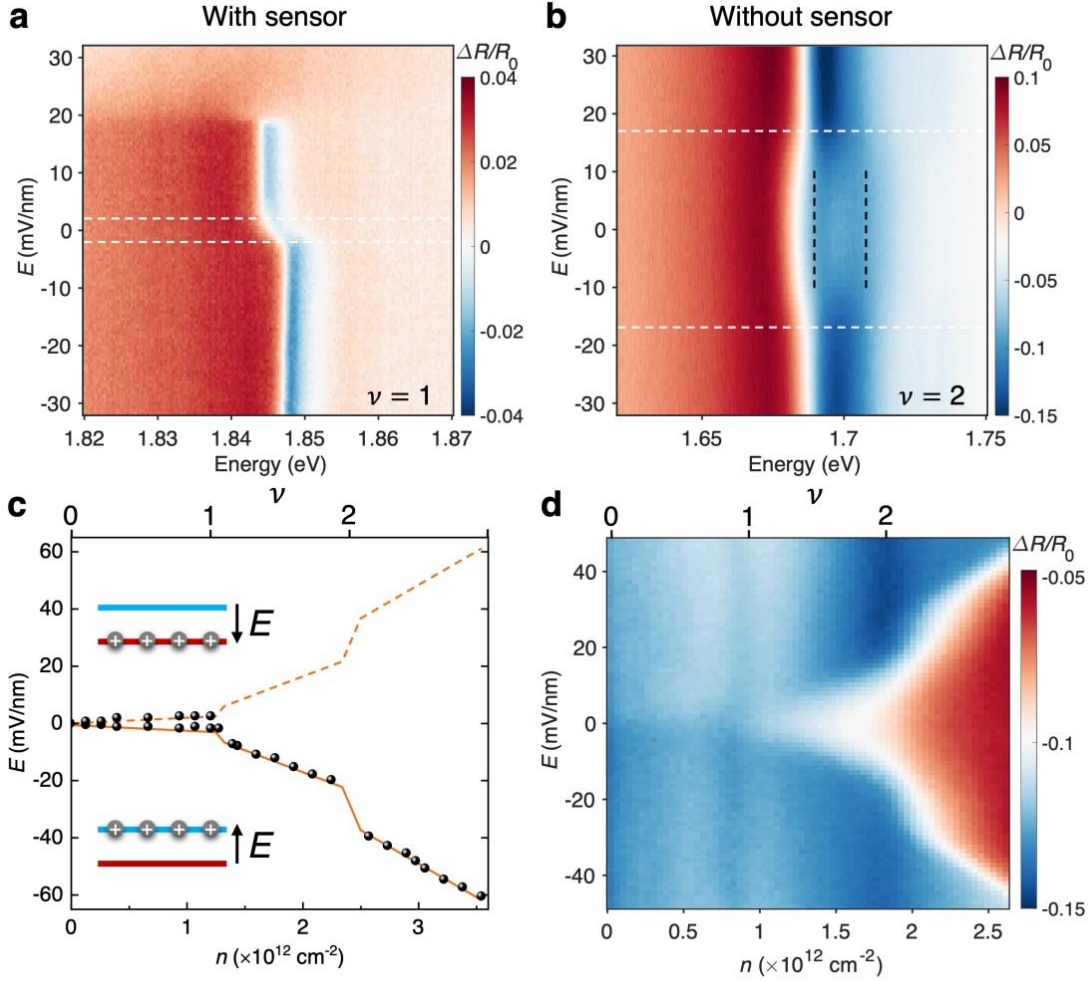

**Figure 2 | Electric-field controlled layer polarization. a, b,** Electric-field dependence of the reflectance contrast spectrum of the sensor 2s exciton at $\nu = 1$ (**a**) and of the moiré exciton of the twisted bilayer at $\nu = 2$ (**b**). The 2s resonance energy shifts near $E = 0$ V/nm due to charge transfer within the moiré bilayer at $\nu = 1$. The layer polarization is saturated beyond $E_c$ (white dashed lines). Above 20 mV/nm the sensor is doped and the 2s exciton is quenched. At $\nu = 2$, two moiré exciton features (guided by the black dashed lines) are replaced by one prominent feature above $E_c$ (white dashed lines). **c,** Doping dependence of $E_c$ (symbols) determined by the exciton sensor. The solid line is a guide to the eye. The dashed line is obtained by inverting the solid line with respect to $E = 0$. The insets illustrate the layer polarization beyond $E_c$. **d,** Averaged reflectance contrast over 5 meV near the moiré exciton resonance of the twisted bilayer at 1.70 eV. The strong and weak contrast regions correspond to layer polarization $|P| = 1$ and $< 1$, respectively. The critical field $E_c$ in **c** is slightly larger than in **d** likely due to the larger twist angle (by about 0.1°).

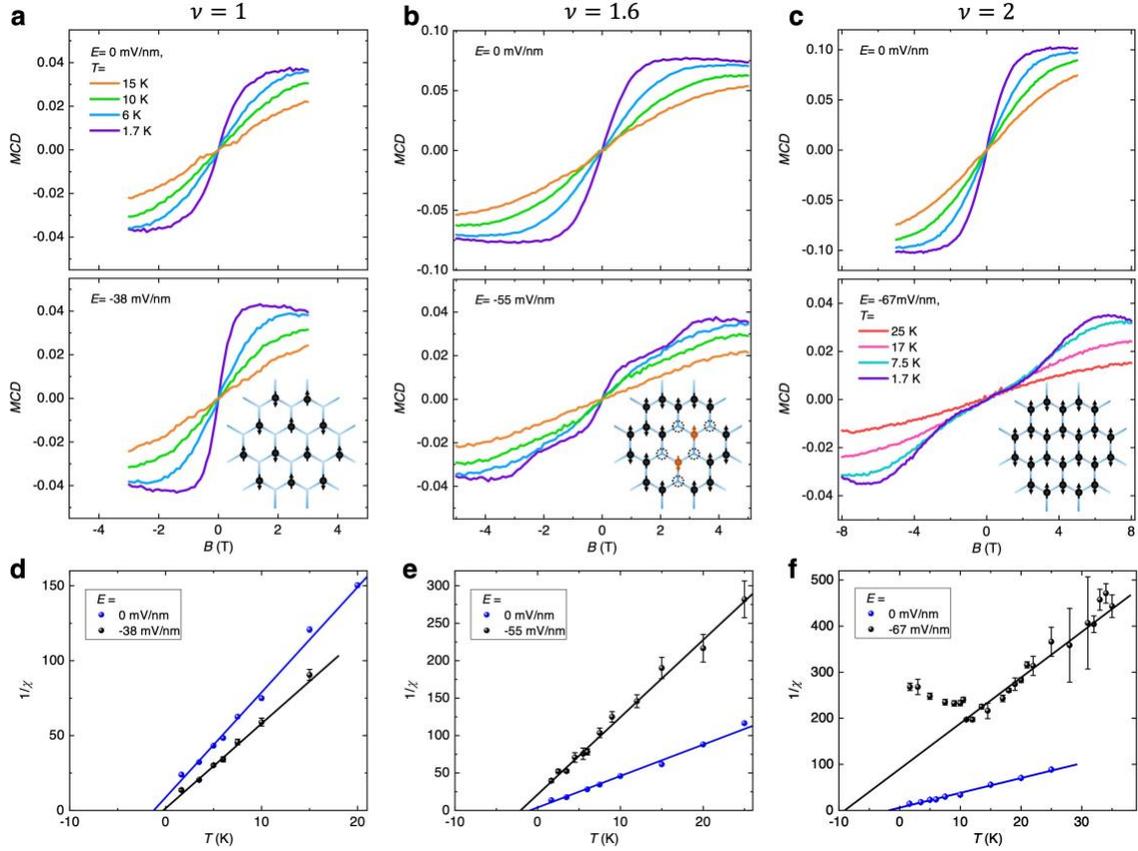

**Figure 3 | Magnetic properties. a-c,** Magnetic-field dependence of MCD at representative temperatures for $\nu = 1$ (**a**), $\nu = 1.6$ (**b**) and $\nu = 2$ (**c**). The top and bottom panels are for $P = 0$ and $P = 1$, respectively. The legend in **a** defines the temperature in all panels except the bottom panel of **c**. The response is PM for all cases with $P = 0$. With $P = 1$, the response is PM for $\nu = 1$ and metamagnetic for $\nu = 2$. For $1 < \nu < 2$ both responses are present with their ratio depending on doping density. The insets illustrate the charge/spin configuration with $P = 1$ at zero magnetic field and 1.7 K. Charges are localized on the MX site (forming a triangular lattice) with random spin orientations at $\nu = 1$. Charges are segregated into clusters that are AF coupled (black arrows) and isolated spins (orange arrows) at $\nu = 1.6$. The dotted circles denote empty sites. Charges form a honeycomb lattice with Néel type AF order at $\nu = 2$. **d-f,** Temperature dependence of the inverse magnetic susceptibility for $\nu = 1$ (**d**), $\nu = 1.6$ (**e**) and $\nu = 2$ (**f**). Symbols: experiment; lines: fit to the Curie-Weiss law. Distinct magnetic responses are observed for the $P = 0$ (blue) and the $P = 1$ case (black).

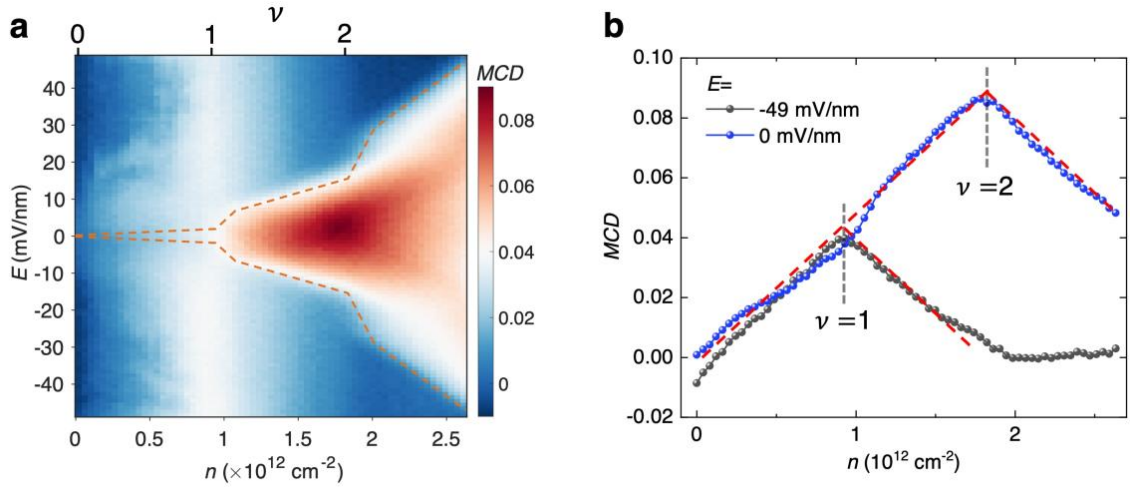

**Figure 4 | Local moments and antiferromagnetic clusters. a,** MCD at $B = 2$ T as a function of electric field and doping density/filling factor. The MCD is probed at a wavelength near the moiré exciton resonance of the twisted bilayer. It is approximately proportional to the density of nearly isolated local moments. The dashed lines show the threshold electric field determined from Fig. 2. Large local moment density is observed for the layer-unpolarized state. **b,** Linecuts of **a** showing doping dependence of MCD with $P = 0$ (blue) and $P = 1$ (black). The red dashed lines denote the expected density dependence of the local moment density for both cases.

**Extended Data Figures**

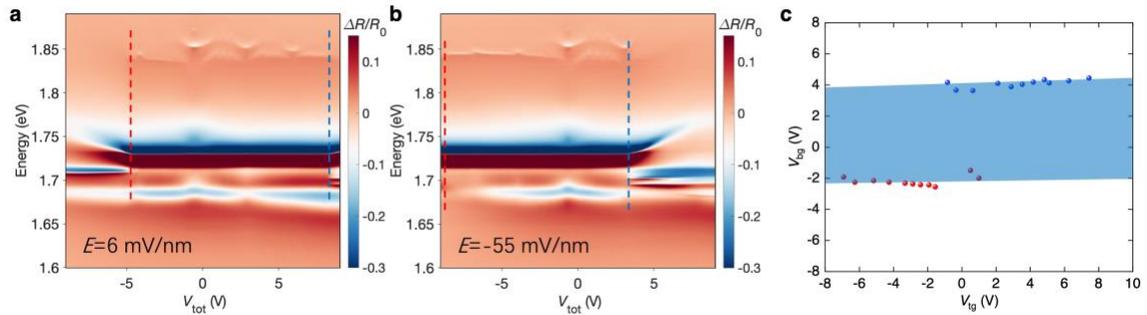

**Extended Data Figure 1 | Operating gate-voltage range of the sensor. a,b**, Gate dependence of the reflectance contrast spectrum of the device with the sensor at $E = 6$ mV/nm (**a**) and $-55$ mV/nm (**b**). The gate voltage $V_{tot}$ (= $V_{tg} + V_{bg}$) is proportional to the total charge density in the moiré bilayer and the sensor layer. The red and blue dashed lines mark the limits of gate voltage beyond which the sensor becomes doped, as evidenced by the emergence of the polarons (1.7 – 1.75 eV) and disappearance of the 2s state (~ 1.85 eV). **c**, Operating range of the gate voltage (shaded area). It is primarily determined by the back gate voltage (-2 V, 4V). The sensor is closer to the back gate. The blue and red symbols are determined from the boundaries in **a**, **b** and similar measurements at other electric fields.

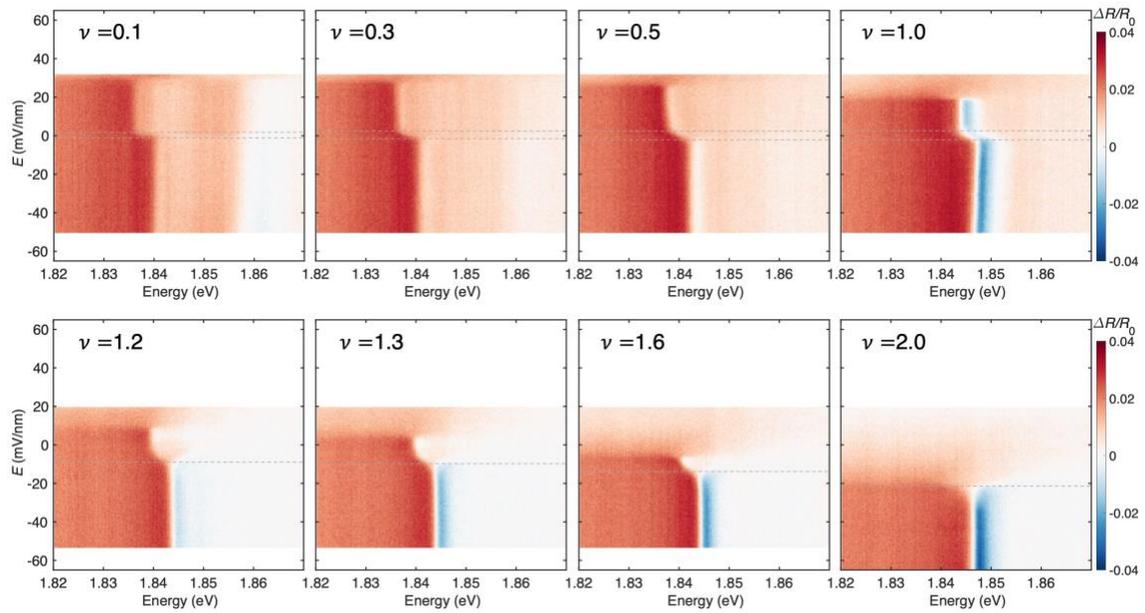

**Extended Data Figure 2 | Layer polarization probed by the sensor 2s exciton.** The reflectance contrast spectrum of the sensor 2s exciton as a function of electric field at representative doping densities. The case of $\nu = 1$ is shown in Fig. 2a of the main text. The gray dashed lines denote the critical electric fields $E_c$.

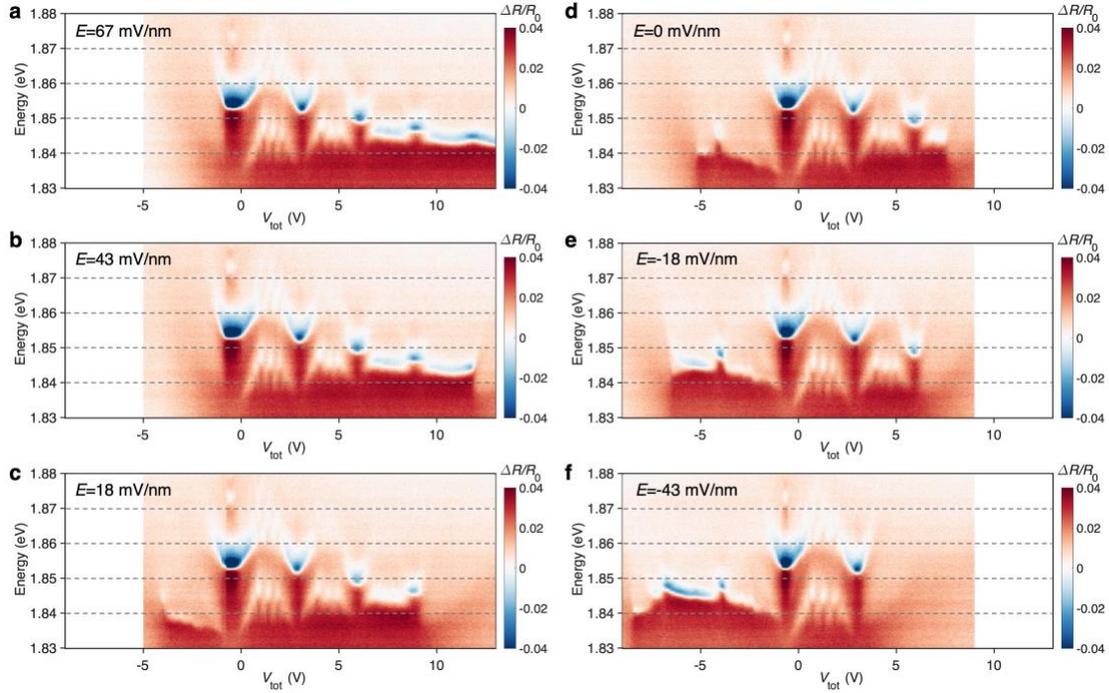

**Extended Data Figure 3 | Electric-field effect on the insulating states. a-f**, Gate dependence of the reflectance contrast spectrum of the sensor 2s exciton at representative electric fields, including 67 mV/nm (**a**), 43 mV/nm (**b**), 18 mV/nm (**c**), 0 mV/nm (**d**), -18 mV/nm (**e**), and -43 mV/nm (**f**). Electrons and holes are introduced into the channel for $V_{tot}$ above and below about -1 V, respectively. The electric field limits the operating voltage range of the sensor. In addition, notable 2s exciton energy shift is observed on the hole side, but not on the electron side, when the electric field switches sign. The result indicates strong layer hybridization for the conduction bands in twisted $WSe_2$, the band edge of which is located at the Q point of the Brillouin zone.

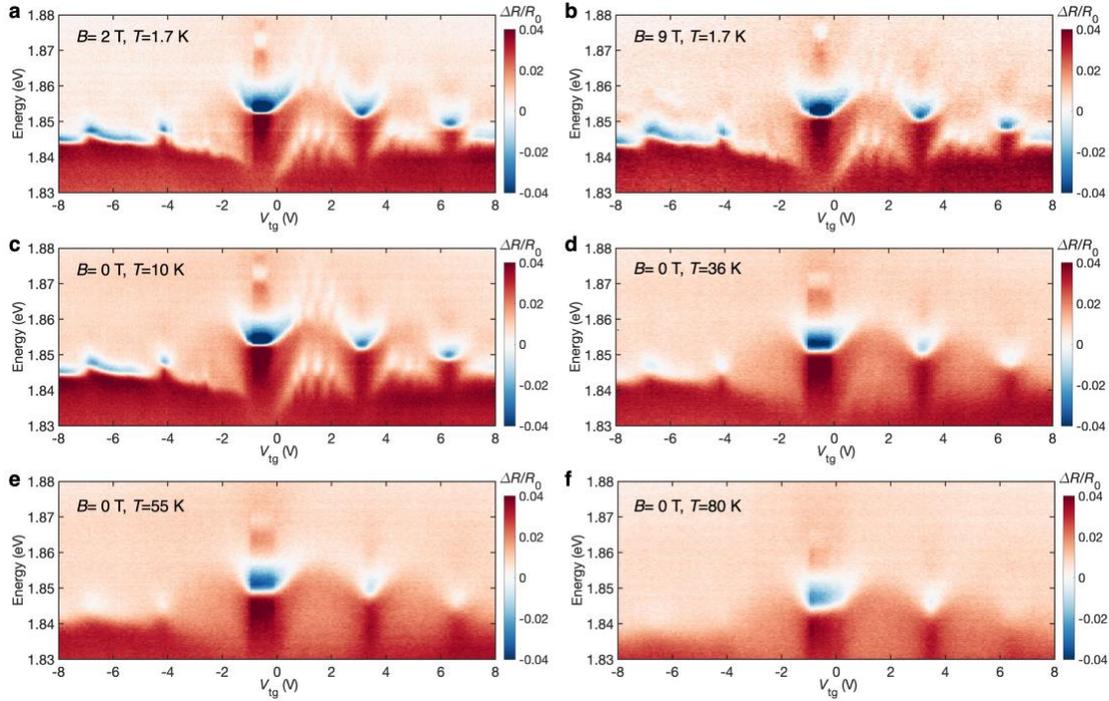

**Extended Data Figure 4 | Temperature and magnetic-field effects on the insulating states. a-f**, Top gate dependence of the reflectance contrast spectrum of the sensor 2s exciton at 1.7 K and under magnetic field of 2 T (**a**) and 9 T (**b**), as well as under zero magnetic field at 10 K (**c**), 36 K (**d**), 55 K (**e**) and 80 K (**f**). The back gate is fixed at 0 V. The corresponding electric field varies for different doping densities. Particularly, the electric field is -43mV/nm for $\nu = 1$ and -76mV/nm for $\nu = 2$ on the hole side, under both of which the moiré bilayer is fully layer-polarized. The correlated insulating states for holes at $\nu = 1$ and 2 are robust against magnetic field for the entire accessible range (9 T) and against thermal melting up to about 80 K.

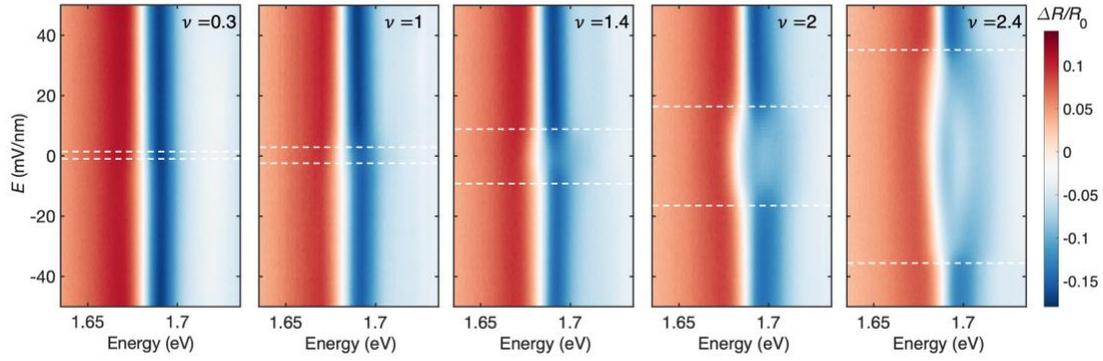

**Extended Data Figure 5 | Layer polarization probed by the moiré exciton.** Reflectance contrast spectrum of the moiré exciton as a function of electric field at representative hole filling factors. The result for $\nu = 2$ is shown in Fig. 2b of the main text. The spectra are measured from the region without the sensor. The white dashed lines mark $E_c$.

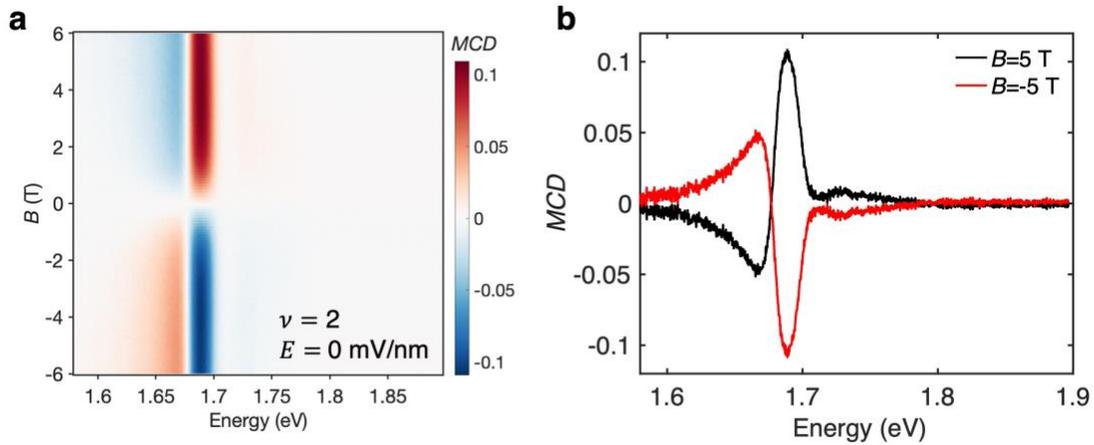

**Extended Data Figure 6 | MCD measurement. a**, Magnetic-field dependence of the MCD spectrum at $\nu = 2$ under zero electric field. **b**, Linecuts of **a** at $B = 5$ T (black line) and -5 T (red line). The MCD is determined as $\frac{R_+(B)-R_+(-B)}{R_+(B)+R_+(-B)}$, where $R_+(B)$ is the reflectance contrast of left circular polarized light under magnetic field $B$.

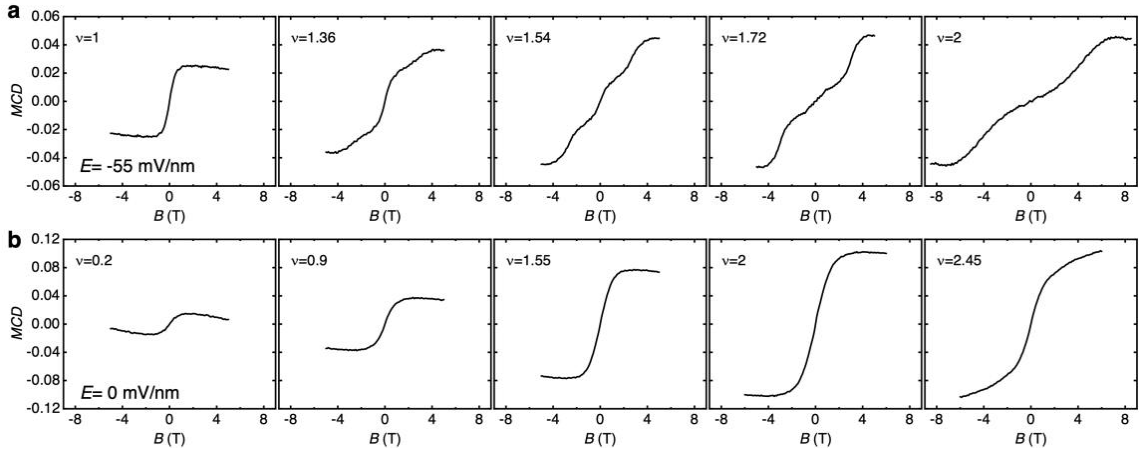

**Extended Data Figure 7 | MCD at different doping densities.** Magnetic-field dependence of MCD for layer-polarized ($E$ = -55 mV/nm) (**a**) and layer-unpolarized holes ($E$ = 0 mV/nm) (**b**) at representative filling factors. The MCD is obtained by averaging the MCD spectrum over a spectral window of 5 meV centered at the exciton resonance (≈ 1.69 eV).

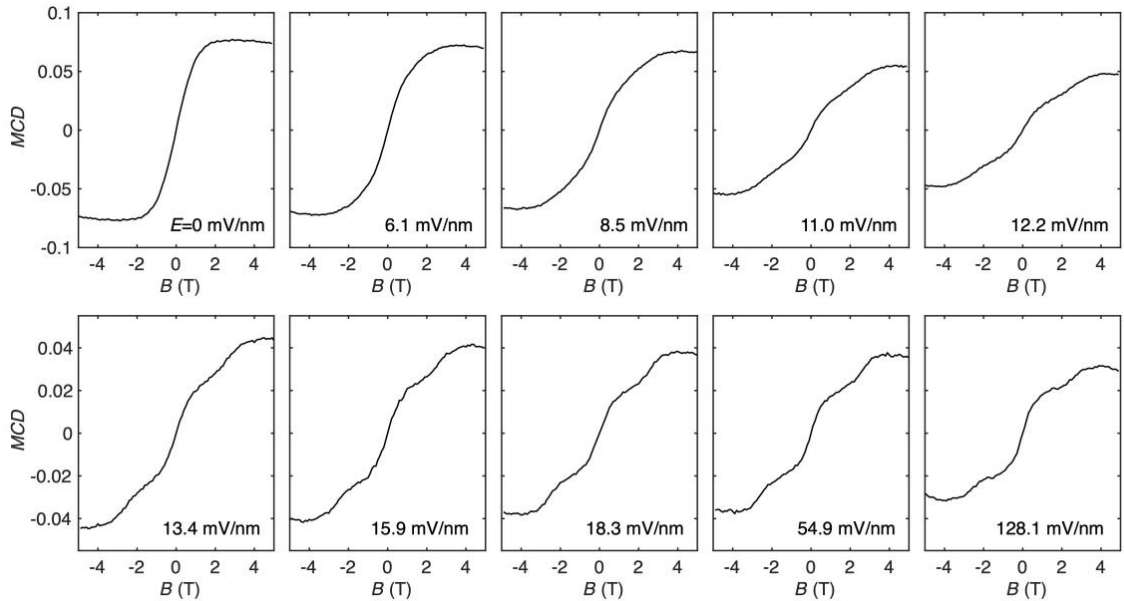

**Extended Data Figure 8 | MCD at $\nu = 1.6$ under different electric fields.** Magnetic-field dependence of MCD under electric field ranging from 0 to 128 mV/nm. The MCD is obtained by averaging the MCD spectrum over a spectral window of 5 meV centered at the exciton resonance (≈ 1.69 eV). The response is PM at small electric fields. A metamagnetic component emerges above about 8.5 mV/nm, which agrees well with the measured threshold electric field $E_c$ for full layer polarization.

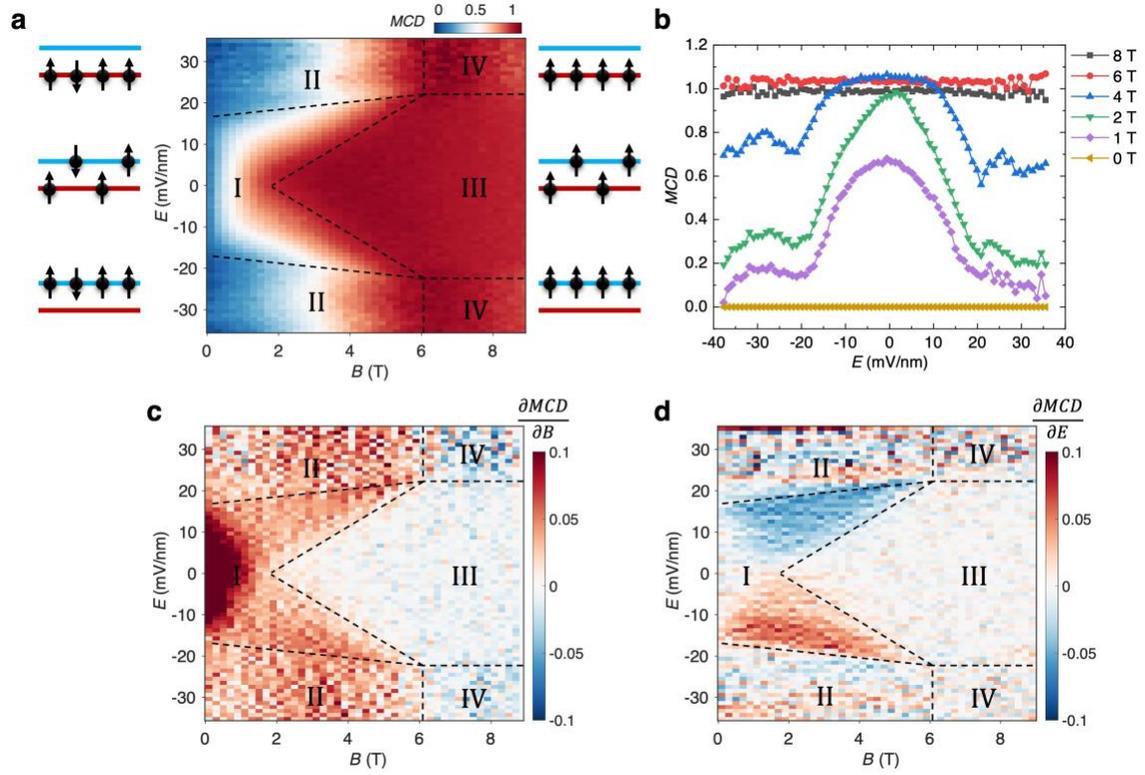

**Extended Data Figure 9 | Magneto-electric (ME) effect at $\nu = 2$. a,c,d,** Normalized MCD (by its value at 9 T) (**a**) and its derivative with respect to $B$ (**c**) and to $E$ (**d**) as a function of the electric and magnetic fields. The four regions (defined by the dashed lines) correspond to: (I) $|M|, |P| < 1$; (II) $|M| < 1, |P| = 1$; (III) $|M| = 1, |P| < 1$; and (IV) $|M|, |P| = 1$. The boundaries are identified by the derivatives in **c** and **d**. The insets in **a** on the left and right illustrate, respectively, the spin and charge configuration at small magnetic fields and after magnetic saturation. On each side the three insets correspond to $P = -1, 0$ and $1$ from top to bottom. **b**, Vertical linecuts of **a** at representative magnetic fields. Strong ME effect is observed for the intermediate magnetic fields, under which the electric field changes the magnetization by over 60%.

21